\documentclass{kluwer}

\newdisplay{guess}{Conjecture}
\input psfig.tex
\def\ie{{\it i.e.}\ }

\def\viz{{\it viz.}\ }

\begin{document}
\begin{article}
\begin{opening}     
\title{Statistical study of short quiescent times between flares}
\author{S\'ebastien \surname{Galtier}}
\runningauthor{S\'ebastien Galtier}
\runningtitle{Statistical study of short quiescent times between flares}
\institute{Mathematics Institute, University of Warwick, Coventry, CV4 7AL, U.K.}

\begin{abstract}
The study of the statistical distribution of short quiescent times ($\le$ 30 
minutes) between solar flares has been investigated on a 1D MHD model. A power 
law behaviour is found which indicates the existence of sympathetic flaring. 
This prediction is supported by recent observations. 
\end{abstract}
\end{opening}


Solar flares are the most violent phenomena observed in the solar system. 
It is believed that they play a fundamental role in maintaining a ``hot'' 
corona. Statistical analyses of observational data 
\cite{Dennis85,Crosby93,Pearce93}, in terms of probability 
distribution functions (pdfs), revealed the scale-invariance property of 
solar flares: the pdfs $N(x)$ of different quantities $x$, such as the 
peak luminosity (P), the total energy (E) or the duration (D) of flare events,
follow widely extended power laws, \viz $N(x) \sim x^{-\alpha_x}$ 
$(x=\hbox{P, E or D)}$ with the indices $\alpha_P \approx 1.7$, 
$\alpha_E \approx 1.5$ and  $\alpha_D \approx 2.2$. This apparent absence of 
a characteristic length scale 
allowed us to see solar flares in a new way, like a Self-Organised Critical 
(SOC) system \cite{Lu91,Vlahos95}. Although SOC models do not make a direct 
connection with physical ingredients, like the magnetic field in the case of 
flares, with simple rules they are able to reproduce well the statistical 
properties observed. The limitation of SOC models in the predictability of 
flares statistics was however recently discussed by \citeauthor{Boffetta} 
(\citeyear{Boffetta}). The authors have studied the pdf of long quiescent 
times  (also called waiting times), $\tau$ (for $\tau \ge 6$ hours), defined 
as the time intervals between two successive bursts. This observational analysis, 
based on 20 years of data obtained from the GOES sensors shows clearly a power law 
behaviour with an index $\alpha_{\tau} \approx 2.4$. 
This kind of behaviour is an indication of the existence of sympathetic flaring 
(\ie the triggering of one flare by another) on long time-scales. It corroborates 
what we already knew for short quiescent times \cite{Pearce93,Wheatland}. Then 
\citeauthor{Boffetta} (\citeyear{Boffetta}) concluded that this result was in
contradiction with SOC models of solar flares, where the system is driven at a 
constant rate \cite{Lu91}, and thus for which events occur randomly in time as 
a Poisson process. Note that there are many SOC models in the literature like 
nonconservative models \cite{chris92} which are able to produce power laws for 
the distribution of quiescent times. Therefore the distinction discussed above 
between SOC and turbulence is not so obvious (although it is for many other aspects!).

Do we really have observed a signature of sympathetic flaring at any time-scale~? 
According to \citeauthor{Wheatland00} (\citeyear{Wheatland00}) the answer is no. 
Indeed in his paper it is shown that the apparent power law behaviour observed for 
long quiescent times is simply the consequence of the wide variation in the flaring 
rate during several solar cycles. However for 
short quiescent times we do have sympathetic flaring. The overabundance of short 
quiescent times (10~s to 10 minutes) constructed by \citeauthor{Wheatland} 
(\citeyear{Wheatland}) from bursts observed in hard X-ray may be interpreted as a
time correlation between secondary events (bursts) generated by a major event 
(flare). Then solar flares, defined as events having several episodes of hard 
X-ray emission, would be independent events. 

A first investigation of the quiescent times statistics in the context of 
magnetohydrodynamics (MHD) is described by \citeauthor{Boffetta} 
(\citeyear{Boffetta}) through a shell model. The idea of a shell model is to 
reproduce the non-linear dynamics by including only a set of nearest-neighbour 
non-linear interactions while preserving the quadratic invariants, but otherwise 
ignoring the details of the spatial structures. In spite of its simplicity, this 
model appears to be rich enough to reproduce the observed statistics for P, D 
and E (see above). However, and in apparent contradiction with the observations, 
a power law behaviour is also observed for the pdf of the long quiescent times. 

The aim of this paper is to investigate the statistical distribution of short 
quiescent times ($\le$ 30 minutes) on a 1D MHD model of coronal loops 
\cite{Galtier1}. 
Several turbulent MHD models of coronal loops have been recently developed but few 
statistical studies of dissipative events have been reported in the literature 
mainly because of the difficulty, from a numerical point of view, to produce 
statistical data of good quality \cite{Einaudi96,dmitruk98,George98b}. 
For that reason, a reduction to a 1D MHD model has been proposed where the 
magnetic loop at the origin of the flare is 
reduced to a magnetic line. Then the events generated are shocks whose 
dissipation is a source of heating for the corona. It was shown that this model 
was able to reproduce solar flares properties like the histogram of peak 
luminosity \cite{Galtier1,Galtier2}. 
\begin{figure}
\centerline{\hbox{\psfig{figure=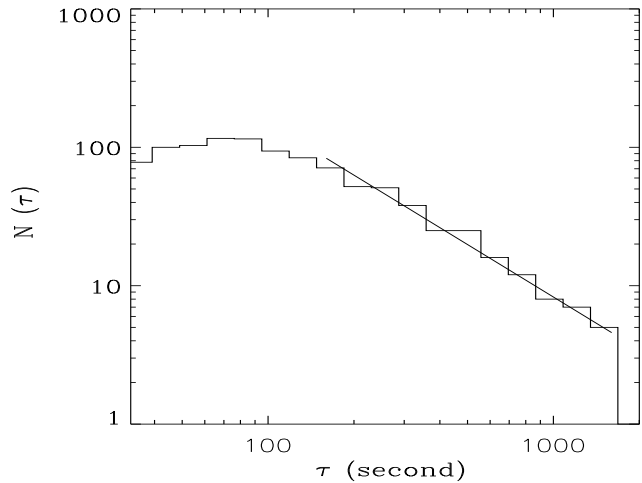,bbllx=.5cm,bblly=.5cm,height=7cm,width=9.5cm}}}
\caption{Histogram $N(\tau)$ of the short quiescent times (in second) between two 
successive dissipative events. The straight line corresponds to a slope of $-2.3$.}
\label{fig1}
\end{figure}
The details and the results of the numerical computation considered here, with a 
spatial resolution of $4,096$ grid points, are given in \citeauthor{Galtier2} 
(\citeyear{Galtier2}). 
The main signal analysed in the present study is the Joule dissipation 
($\sim \langle J^2(x) \rangle$, where ${\bf J}$ is the current density) averaged 
over the length of the loop for a simulation of 42 hours of solar activity. Each 
dissipative event of the signal defines a flare and the quiescent times are then 
taken to be the time intervals between two successive events, \ie two successive 
maxima, following the criterion also used in \citeauthor{Wheatland} 
(\citeyear{Wheatland}) and in \citeauthor{Boffetta} (\citeyear{Boffetta}). The 
computation of the histogram of the quiescent times is displayed in Figure 1 in
log-log coordinates; the unit time chosen is the second. A clear power law 
behaviour is revealed over a range of time from 3 minutes to 30 minutes, with 
an index $\alpha_{\tau} \approx 2.3$. It indicates the existence of sympathetic 
flaring in agreement with the observations made by \citeauthor{Pearce93} 
(\citeyear{Pearce93}) for a range of short quiescent times from 1 to 60 minutes, 
and with the observations made by \citeauthor{Wheatland} (\citeyear{Wheatland}) 
where an overabundance of short quiescent times is seen from 10~s to 10 minutes. 
In fact in the latter case the data can also be fitted with a power from $10^2$s 
to $10^4$s. We have to emphasise here that, (a) the power law index obtained with 
the MHD simulation is different from the observations. However it is claimed that 
the most important result is the power law behaviour which implies the existence 
of sympathetic flaring for short time-scales. (b) Some precautions have to be taken
with the result obtained by \citeauthor{Pearce93} (\citeyear{Pearce93}) since the
authors did not take into account the variation in the flaring rate like in 
\citeauthor{Wheatland} (\citeyear{Wheatland}). 

The physical picture for long and short quiescent times appears to be very 
different. For long quiescent times (longer than the lifetime of a flare) no time 
correlation seems to be observed, however for short quiescent times (of the order 
or smaller than the lifetime of a flare) we do observe a time correlation. As 
it is proposed by \citeauthor{Wheatland} (\citeyear{Wheatland}) the origin of
sympathetic flaring in the MHD model is possibly due to the correlation 
between secondary events (bursts) generated by a major event (flare). In fact 
typical dissipative events last several minutes \cite{Galtier2} and are often 
composed in a main flare followed by secondary events. The detailed analysis, 
including spatial information along the coronal loop, on typical examples of 
dissipative events reveals indeed a non-trivial correlation between bursts 
\cite{Galtier3} for short time-scales. This spatial information available in 
the MHD model appears to be an important ingredient to understand the origin
of sympathetic flaring.

\acknowledgements
I am thankful to N. Watkins and A. Pouquet for valuable discussions. Grant from 
EEC (contract FMRX-CT98-0175) and from CNRS (PNST) are gratefully acknowledged. 

\end{article}

\begin{thebibliography}{}
%
\bibitem[\protect\citeauthoryear{Boffetta et al.}{1999}]{Boffetta}
G. Boffetta, V. Carbone, P. Giuliani, P. Veltri and A. Vulpiani. 
\newblock {Power laws in solar flares: SOC or turbulence\,?}
\newblock {\em Phys. Rev. Lett.}, 83: 4662--4665, 1999.
%
\bibitem[\protect\citeauthoryear{Christensen and Olami}{1992}]{chris92}
K. Christensen and Z. Olami.
\newblock {Variation of the Gutenberg-Richter b value and nontrivial temporal
correlation in a spring-block model for earthquakes}
\newblock {\em J. Geophys. Res.}, 97: 8729--8735, 1992.
%
\bibitem[\protect\citeauthoryear{Crosby et al.}{1993}]{Crosby93}
N.~B. Crosby, M.~J. Aschwanden and B.~R. Dennis. 
\newblock {Frequency distributions and correlations of solar X--ray flare parameters}.
\newblock {\em Solar Phys.}, 143: 275--299, 1993.
%
\bibitem[\protect\citeauthoryear{Dennis}{1985}]{Dennis85}
B.~R. Dennis. 
\newblock {Solar hard X--ray}.
\newblock {\em Solar Phys.}, 100: 465--490, 1985.
%
\bibitem[\protect\citeauthoryear{Dmitruk et al.}{1998}]{dmitruk98}
P. Dmitruk, D.~O. G\'omez and E.~E. DeLuca, 
\newblock {MHD turbulence of coronal active regions and the distribution of nanoflares}.
\newblock {\em Astrophys. J.}, 505: 974--983, 1998.
%
\bibitem[\protect\citeauthoryear{Einaudi et al.}{1996}]{Einaudi96}
G. Einaudi, M. Velli, H. Politano and A. Pouquet. 
\newblock {Energy release in a turbulent corona}.
\newblock {\em Astrophys. J.}, 457: L113--L116, 1996.
%
\bibitem[\protect\citeauthoryear{Galtier and Pouquet}{1998}]{Galtier1}
S. Galtier and A. Pouquet.
\newblock {Solar flares statistics with a one-dimensional MHD model}.
\newblock {\em Solar Phys.}, 179: 141--165, 1998.
%
\bibitem[\protect\citeauthoryear{Galtier}{1999}]{Galtier2}
S. Galtier.
\newblock {A 1D MHD model of solar flares: emergence of a population of weak~events
and a possible road towards nanoflares}.
\newblock {\em Astrophys. J.}, 521: 483--489, 1999.
%
\bibitem[\protect\citeauthoryear{Galtier}{2000}]{Galtier3}
S. Galtier.
\newblock {A classification of solar flares stemming from a 1D MHD model}.
\newblock {In preparation}, 2000.
%
\bibitem[\protect\citeauthoryear{Georgoulis et al.}{1998}]{George98b}
M.~K Georgoulis, M. Velli and G. Einaudi.
\newblock {Statistical properties of magnetic activity in the solar corona}.
\newblock {\em Astrophys. J.}, 497: 957--966, 1998.
%
\bibitem[\protect\citeauthoryear{Lu and Hamilton}{1991}]{Lu91}
E.~T. Lu and R.~J. Hamilton.
\newblock {Avalanches and the distribution of solar flares}.
\newblock {\em Astrophys. J.}, 380: L89--L92, 1991. 
%
\bibitem[\protect\citeauthoryear{Pearce et al.}{1993}]{Pearce93}
G. Pearce, A.~K. Rowe and J. Yeung. 
\newblock {A statistical analysis of hard X--ray solar flares}.
\newblock {\em Astrophys. Space Sci.}, 208: 99--111, 1993.
%
\bibitem[\protect\citeauthoryear{Vlahos et al.}{1995}]{Vlahos95}
L. Vlahos, M. Georgoulis, R. Kluiving and P. Paschos.
\newblock {The statistical flare}.
\newblock {\em Astron. Astrophys.}, 299: 897--911, 1995.
%
\bibitem[\protect\citeauthoryear{Wheatland et al.}{1998}]{Wheatland}
M.~S. Wheatland, P.~A. Sturrock and J.~M. McTiernan. 
\newblock {The waiting-time~distri\-bution of solar flare hard X--ray bursts}.
\newblock {\em Astrophys. J.}, 509: 448--455, 1998. 
%
\bibitem[\protect\citeauthoryear{Wheatland}{2000}]{Wheatland00}
M.~S. Wheatland. 
\newblock {The origin of the solar flare waiting-time distribution}.
\newblock {\em Astrophys. J. Lett.}, 536: L109--L112, 2000. 
%
\end{thebibliography}
\end{document}